\documentclass[twocolumn,showpacs,showkeys,superscriptaddress]{revtex4}
\usepackage{amsmath}  
\usepackage{amsfonts} 
\usepackage{graphicx} 
\usepackage{color}
\usepackage{siunitx}
\usepackage{subfig}
\usepackage{blindtext}

\begin{document}

\title{Accelerating solutions to  diffusion equation}

\author{Felipe A. Asenjo}
\email{felipe.asenjo@uai.cl}
\affiliation{Facultad de Ingenier\'ia y Ciencias,
Universidad Adolfo Ib\'a\~nez, Santiago 7491169, Chile.}
\author{Sergio A. Hojman}
\email{sergio.hojman@uai.cl}
\affiliation{Departamento de Ciencias, Facultad de Artes Liberales,
Universidad Adolfo Ib\'a\~nez, Santiago 7491169, Chile.}
\affiliation{Departamento de F\'{\i}sica, Facultad de Ciencias, Universidad de Chile,
Santiago 7800003, Chile.}
\affiliation{Centro de Recursos Educativos Avanzados,
CREA, Santiago 7500018, Chile.}

\begin{abstract}
We report accelerating diffusive solutions to the diffusion equation with a constant diffusion tensor. The maximum values of the diffusion density evolve in an accelerating fashion described by Airy functions. We show the diffusive accelerating behavior for one--dimensional systems, as well as for a  general three--dimensional case.
 We also construct a modulated modified form of the diffusion solution  that retains the accelerating features.
\end{abstract}


\maketitle
 
\section{Introduction}

Diffusion is one of the most important problems in physics. Its applications include random walk and heat diffusion \cite{crank}, economy \cite{blackscholes}
and biophysics \cite{bioph}, for instance.  In the most general fashion, the diffusion equation for a three--dimensional diffusing material with density  $\phi(t,{\bf x})$, 
can be written as
\begin{equation}
  \frac{\partial \phi}{\partial t}=\frac{\partial}{\partial x_i}\left(D^{ij}\frac{\partial\phi}{\partial x_j}\right)\, ,
\label{difGenera}
\end{equation}
where $D^{ij}=D^{ij}({\bf x})$ is the symmetric diffusion tensor, which in principle depends on the position ${\bf x}$ where diffusion takes place. For Eq.~\eqref{difGenera} we have used the Einstein summation convention for $i, j=1, 2, 3$.

Along this work, we show that  accelerating diffusive solutions, described in terms of Airy functions,
are possible for a general case of Eq.~\eqref{difGenera} when the  diffusion tensor $D^{ij}$ has  constant components. 
The  solutions we present have accelerating properties such that the diffusion evolution of the local density
has curved trajectories in a space-time plane. In that sense, the position $x_M$ of the maxima for the diffusive density accelerate, i.e, $d^2 x_M/dt^2\neq 0$, in such a way that the density of the diffusing material has a diffusion which is different from $\sim\sqrt{t}$, meaning that it differs from standard solutions. 

The aim of this work is to present these accelerating solutions as robust phenomena, which have not been fully explored in diffusion.
We start studying the accelerating behavior for one--dimensional systems, to later show the accelerating solution for a general case.

\section{One--dimensional  diffusion}

\begin{figure*}[ht]
\begin{tabular}{cc}
  \includegraphics[height=75mm]{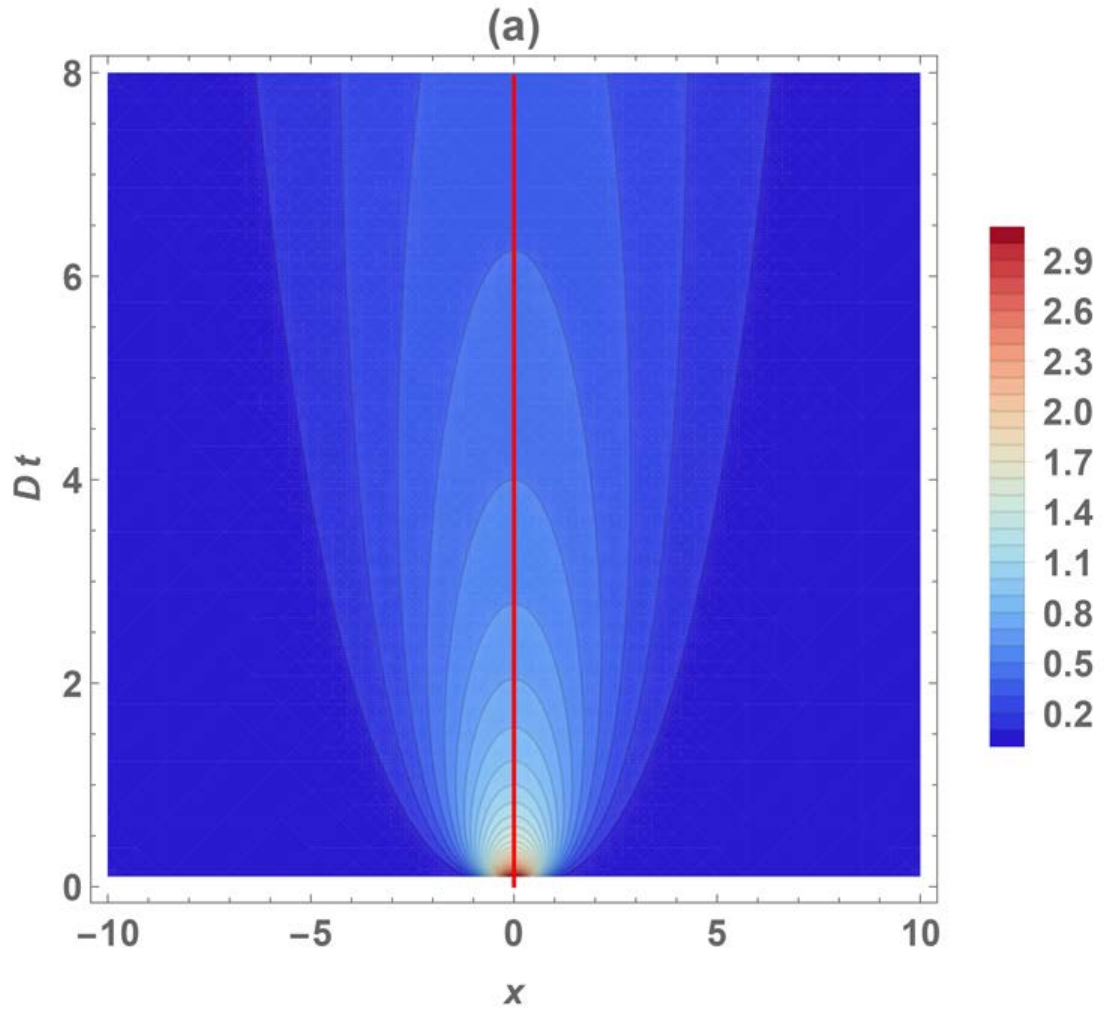} &     \includegraphics[height=75mm]{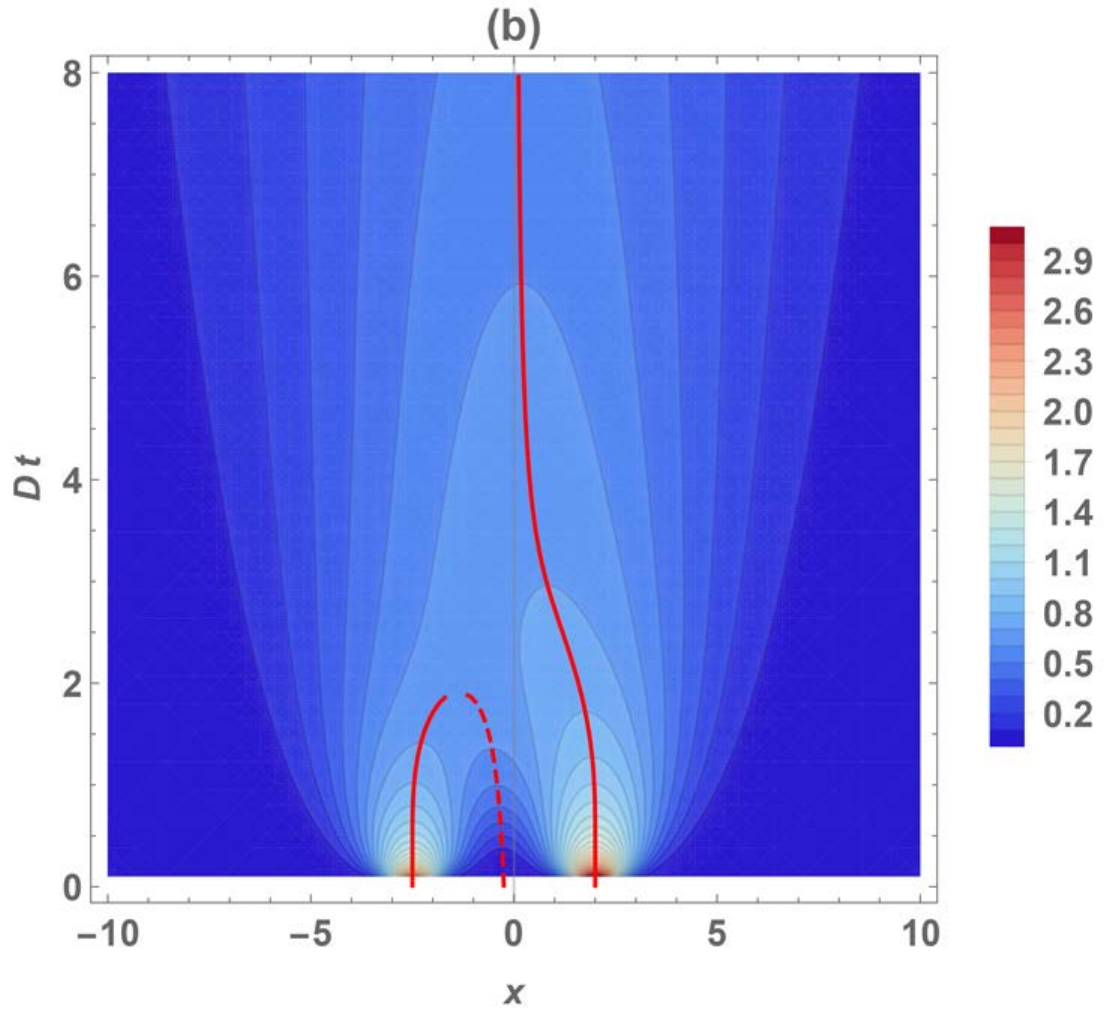} \\
 \includegraphics[height=75mm]{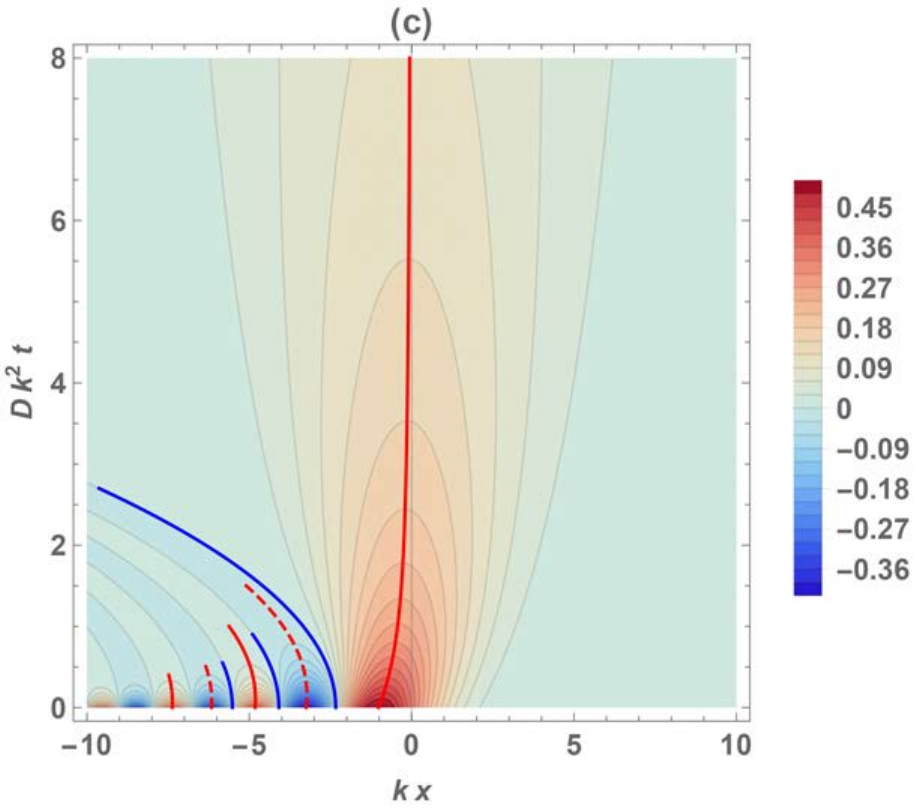} & \includegraphics[height=75mm]{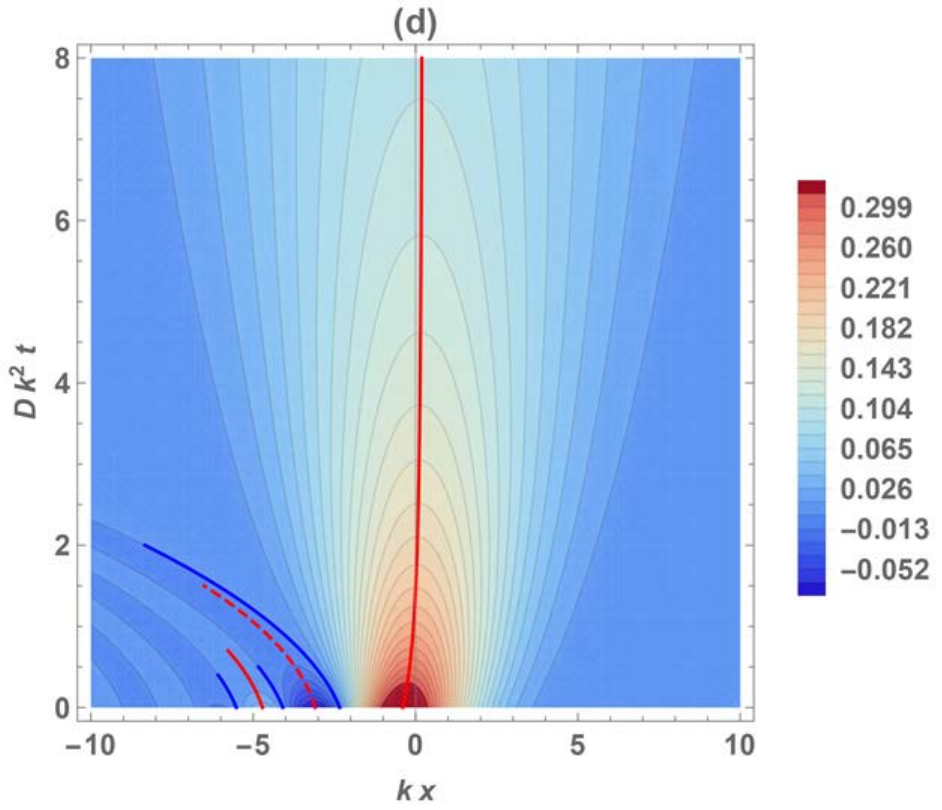} 
\\
\end{tabular}
\caption{Contour plots, in space-time phase space,  showing the evolving dynamics of different diffusion densities. Solid red lines represent the evolution of the position of the maxima of the densities. Dashed red lines represents the evolution of their minima. Blue solid lines  represents the evolution of the zero-level solutions.
 (a) Contour plot   for density \eqref{eqDiclasical}, in terms of variables $x$ and $D\, t$. Its maxima do not move. (b) Contour plot for density \eqref{eqDiclasicaldoble}, in terms of variables $x$ and $D\, t$, showing accelerating properties of their maxima.
(c) Contour plot for density \eqref{airy1}, in terms of variables $k\, x$ and $k^2 D\, t$ (equivalent to density \eqref{airy2} with  $\eta=0$), showing the accelerating evolution of the maxima, minima, and zero-level solutions. (d)  Same for  density \eqref{airy2}, with $\eta=1/2$.
}
\label{fig1final}. 
\end{figure*}

We start simply by studying diffusion in a one--dimensional system with constant diffusive coefficient $D$, such as $D^{ij}=D \delta^{i x}$ along a preferred direction, say $x$.
In this case
a diffusing material with density  $\phi(t,x)$, has a dynamics given by 
\begin{equation}
    \frac{\partial \phi}{\partial t}=D\frac{\partial^2\phi}{\partial x^2}\, ,
    \label{dif}
\end{equation}
which is equivalent to the heat equation (where in such case $\phi$ represents the temperature).
The standard well--known solution of Eq.~\eqref{dif} has the Gaussian form 
\begin{equation}
\phi(t,x)=\frac{1}{\sqrt{D t}}\exp\left(-\frac{x^2}{4D t} \right)\, .
\label{eqDiclasical}
\end{equation}
The diffusive behavior of solution \eqref{eqDiclasical}, in a space-time phase space,
is shown
in Fig.~\ref{fig1final}(a).

 In what follows, we present a different exact solution of Eq.~\eqref{dif} with accelerating properties in the $x-t$ phase space plane.
Several solutions for the nonlinear version of the diffusion equations have been studied (see for example Refs.~\cite{burgers,king1,king2,chole,metzler,chai,said,ander,james} among others), 
and also in quantum systems \cite{tsekov}. However, the simplest linear model described by Eq.~\eqref{dif} has a solution that has not been fully explored, as yet. 
This solution corresponds to an initial value condition for the temperature $\phi(0,x)$ that exhibits an accelerating form of diffusion as it evolves in time.

 Several other known solutions of diffusion have accelerating properties. 
In the context of reaction--difusion equations, different features of initial conditions modify the diffusion behavior \cite{garnier,dstan, hamel,rking}. On the other hand, for instance,
Gaussian--like solutions with multiple local maxima as initial conditions satisfying Eq.~\eqref{dif}, such as
\begin{eqnarray}
\phi(t,x)&=&\frac{1}{\sqrt{D t}}\exp\left(-\frac{(x-x_0)^2}{4D t} \right)\nonumber\\
&&+\frac{\lambda}{\sqrt{D t}}\exp\left(-\frac{(x+x_0/\lambda)^2}{4D t} \right)\, ,
\label{eqDiclasicaldoble}
\end{eqnarray} 
 have accelerating properties. Here, $x_0$ and $\lambda$ are  arbitrary. This solution has the usual diffusive mean square distance $\langle x^2\rangle=\int_{-\infty}^\infty x^2\phi\, dx=4\sqrt{\pi}(1+\lambda)(t+x_0^2/2\lambda)$.
This density is plotted in Fig.~\ref{fig1final}(b), where its maxima are shown in solid red lines, and the minima is displayed in dashed red line.  The trayectories of the maxima approach to $x=0$ for longer times, and they display non-constant acceleration that approaches 0 as time grows. 
This also can be seen in Fig.~\eqref{fig3final}(a).
However this kind of behavior is different from the presented solution in this work, as we will show.

\subsection{Simple accelerating solution}

Let us start with the  following ansatz, containing the simplest form of acceleration $y=x+\beta t^2$ in a one--dimensional system,
\begin{equation}
 \phi(t,x)=f\left(k\, y\right)\exp\left(\zeta(t,x)\right)\, ,
\label{seolprelimial}
\end{equation}
where $f$ and $\zeta$ are functions to be determined, $\beta$ is a constant and 
$k$ is an arbitrary constant with units of inverse length. A simple solution can be found by assuming that $\zeta(t,x)=\alpha\, t\, x+\mu(t)$, in terms of a constant $\alpha$ and a function $\mu$ that depends on time only. Using this ansatz in Eq.~\eqref{dif}, we find that function $f$ is an Airy function, $\beta=D^2 k^3$, $\alpha=D k^3$, and $\mu(t)=2 D^3 k^6 t^3/3$. In this form, the ansatz \eqref{seolprelimial} acquires the form
\begin{equation}
    \phi(t,x)={\mbox{Ai}}\left(k\, x+ k^4 D^2 t^2 \right)\exp\left(k^3 D\, t\, x+\frac{2}{3}k^6 D^3 t^3 \right)\, ,
    \label{airy1}
\end{equation}
which can be straightforwardly proved to be a solution of Eq.~\eqref{dif}.
The diffusive density \eqref{airy1} can be also obtained by the Wick rotation in time of the Berry--Balazs's nondiffracting wavepacket  solution \cite{berry}
for non--relativistic quantum mechanics.

Solution
\eqref{airy1} presents local diffusion with acceleration in the $x-t$ plane. This can be explicitly seen in Fig.~\ref{fig1final}(c), where the density \eqref{airy1} is plotted in terms of dimensionless variables $k\, x$ and $k^2 D\, t$. 
This diffusion oscillates (because of the Airy function) representing different local temperature gradients which can be set as initial conditions. In Fig.~\ref{fig1final}(c), 
we explicitly show the evolution in time of the local maxima  (solid red lines) and local minima (dashed red lines) values of density \eqref{airy1}. 
The maxima and minima have curved trayectories in $x-t$ plane, displaying the  local acceleration features of this solution. 

\begin{figure}[h]
 \includegraphics[height=50mm]{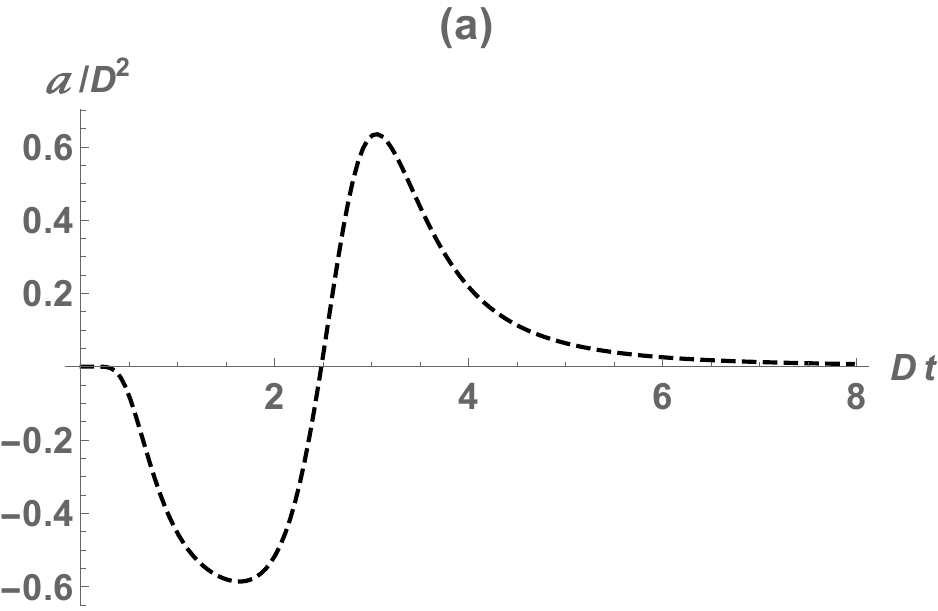}\\
 \includegraphics[height=50mm]{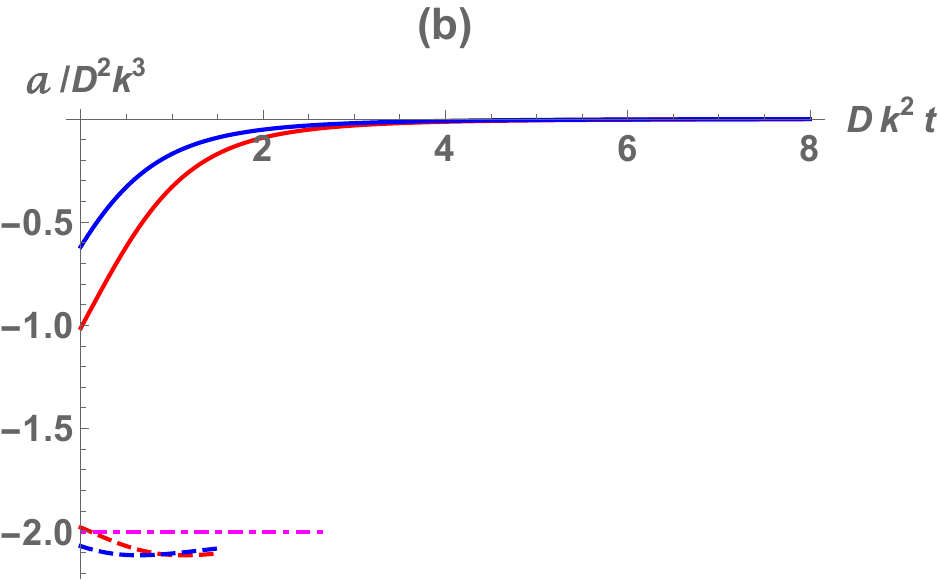}
\caption{Types of acceleration for different local points in previous diffusion solutions. (a) Acceleration $a=d^2 x_M/dt^2$ for maximum maximorum of difussion density \eqref{eqDiclasicaldoble}, depicted in Fig.~\ref{fig1final}(b), as a function of $D t$. (b) Acceleration of  first maxima (solid red line), first minimum (dashed red line), and first zero level (dot-dashed magenta line) of density \eqref{airy1}, as function of $D k^2 t$. Similarly, acceleration of  first maxima (solid blue line), first minimum (dashed blue line), and first zero level (same dot-dashed magenta line) of density \eqref{airy2}, with $\eta=1/2$. }
\label{fig3final}. 
\end{figure}

The dynamics of the local maxima or minima can be found by solving for $x_M(t)$ the following equation
\begin{equation}
\frac{\mbox{dAi}(\xi)}{d\xi}+Dk^2 t\, \mbox{Ai}(\xi)=0\, ,
\end{equation}
where $\xi=k\, x_M+ k^4 D^2 t^2$. This equation is solved numerically and shown in Fig.~\ref{fig1final}(b). Cleary $d x_M/dt\neq 0$ in all cases [different to the Gaussian case of Fig.~\ref{fig1final}(a)]. 
While the maximum maximorum lobe curves towards the positive values of $x$, the other maxima and minima curve to its negative values.
This implies different type of acceleration for different maxima and minima. This can be seen in Fig.~\ref{fig3final}(b),
where $a_M=d^2 x_M/dt^2\neq 0$ is shown for the maximum maximorum (solid red line) and the first minimum (dashed red line) of solution \eqref{airy1}.
 In particular, the main maxima lobe accelerate always with $a_M < 0$, decreasing in magnitude as the time grows. The other maxima and minima have a slightly growing negative acceleration. This explains the different path taken by the main lobe and the other ones.
 Compare all these results with diffusion density \eqref{eqDiclasical} in  Fig.~\ref{fig1final}(a), where its maxima do not accelerate (they follow a straight line trajectory in the $x-t$ plane), or with diffusion \eqref{eqDiclasicaldoble} in Fig.~\ref{fig1final}(b), where the accelerating solutions approaches to $x=0$, changing sign in acceleration in order  
to form a unique Gaussian--like structure with no acceleration.

In addition to this, we can calculate the trajectories of the zero-levels of density \eqref{airy1}. These are the blue lines in Fig.~\ref{fig1final}(c). They display acceleration that can be found analytically, as the zeros of Airy function are known. In this way, for instance, the trajectory of the first zero-level (the first blue line) in Fig.~\ref{fig1final}(c) has a time dependent position $x_z$ given by
\begin{eqnarray}
x_z(t)=-k^3 D^2 t^2+\frac{{\mbox{Ai}}_{z1}}{k}\, ,
\end{eqnarray}
where ${\mbox{Ai}}_{z1}\approx -2.33811$ is the value for the first zero of Airy function [${\mbox{Ai}}({\mbox{Ai}}_{z1})=0$]. It presents a change in position with the acceleration $a_z=d^2x_z/dt^2=-2 k^3 D^2$. This acceleration is shown in Fig.~\ref{fig3final}(b) as the dot-dashed magenta line.
The procedure to obtain the dynamics of other zero-level curves is straightforward.

\begin{figure}[h]
 \includegraphics[height=50mm]{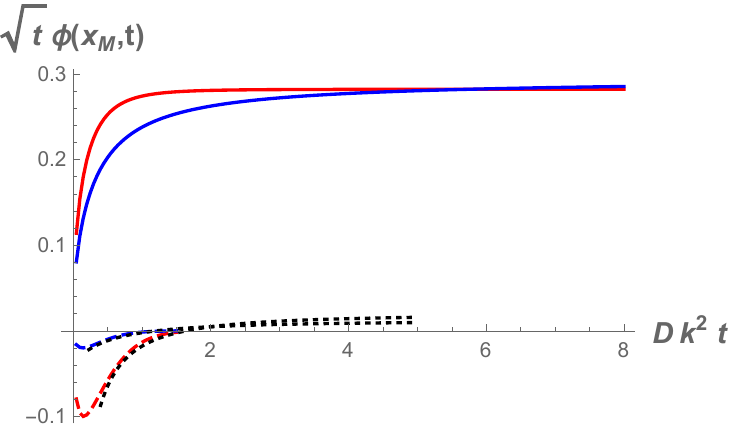}
\caption{Diffusive behavior of accelerating solutions evaluated in maximum maximorum and first minima. Blue solid and dashed lines are for diffusion density \eqref{airy1}. Red solid and dashed lines are for diffusion density \eqref{airy2}.}
\label{fig4}. 
\end{figure}

The asymmetry in the accelerating bearing of local maxima and minima in Fig.\ref{fig3final}(b) can be explained by their asymptotic 
behavior. For the main lobe, the solution \eqref{airy1} decays in $\infty$ as $\sim|\xi|^{-1/4}\exp(-|\xi|^{3/2})$, which is faster than any other local maxima and minima that decay in $-\infty$ as $\sim|\xi|^{-1/4}$  only. These two different kinds of behavior explain also the different local diffusion rates of the solution. We can show that different local solutions of \eqref{airy1} have diffusion rates different to $\sim\sqrt{t}$. Gaussian solutions \eqref{eqDiclasical} and \eqref{eqDiclasicaldoble} have diffusion rates (locally and globally) that tend to  $\sqrt{t}$. In Fig.~\eqref{fig4} we use
density \eqref{airy1} to display $\sqrt{t}\, \phi$
 evaluated in  the maximum maximorum (solid red line) and in the first minimum (dashed redline). This plot indicates that the accelerating solution diffuses in a different way that the Gaussian solution. Its maximum approaches the usual diffusion  $\sim\sqrt{t}$. However, the other maxima and minima show a completely different dynamics, diffusing faster as $\sim t^2$ as time grows (represented as black dotted lines). In general, all local maxima and minima in $(-\infty,0)$ have a $\sim t^2$ diffuse behavior.
 This does not occur in Gaussian-like solutions, and it is a direct consequence of the accelerating properties of the Airy function.

\subsection{Generalized accelerating solution}

Despite of the interesting features of solution \eqref{airy1}, it is possible to construct a modulated/attenuated version of the above solution by an arbitrary parameter. Following  Refs.~\cite{sivi,lekn}, and performing  a Wick rotation in time, the modulated accelerating Airy diffusive solution has  the form
\begin{equation}
    \begin{split}
\phi(t,x)&={\mbox{Ai}}\left(k\, x+ k^4 D^2 t^2+2\eta\, k^2 D\, t \right)\\
&\times\exp\Bigg(k^3 D\, t\, x+\frac{2}{3}k^6 D^3 t^3+\eta k x\\
&\qquad \quad
+2\eta k^4 D^2t^2 +\eta^2 k^2 D\,  t \Bigg)\, , 
\label{airy2}
\end{split}
\end{equation}
which satisfies the diffusion equation \eqref{dif} for an arbitrary dimensionless constant $\eta$. This solution  approaches to a  single profile as $\eta$ increases, but retaining its accelerating features. This can be seen in Fig.~\ref{fig1final}(d), where the density \eqref{airy2} is shown in terms of $k x$ and $k^2 D t$, for
 $\eta=1/2$.

 The accelerating trajectory for the diffusion of the maxima of the density (solid red line) shows that the accelerating behavior is present but reduced.  The same occurs for the minima (dashed red lines) and the zero--level solutions (blue lines). In general, as $\eta$ increases, the acceleration properties  are attenuated.

Similar to the simplest case, the time dependent positions $x_M$ of local maxima and minima of  density \eqref{airy2} can be calculated numerically by solving the equation
\begin{equation}
\frac{\mbox{dAi}\left(\tilde\xi\right)}{d\tilde\xi}+Dk^2 t\, \mbox{Ai}\left(\tilde\xi\right)=0\, ,
\end{equation}
for $x_M$, where now $\tilde\xi=k\, x_M+ k^4 D^2 t^2+2\eta k^2 Dt$. Also, the different zero-levels have time dependent positions $x_{zi}$ (with $i=1,2,3,...$) given by
\begin{eqnarray}
x_{zi}(t)=-k^3 D^2 t^2-2\eta k D t+\frac{{\mbox{Ai}}_{zi}}{k}\, ,
\end{eqnarray}
where ${\mbox{Ai}}_{zi}$ 
are the different zeros of Airy function. All zero-levels anew move with constant acceleration $a_z=-2k^3 D^2$.

The accelerating properties of this modulated diffusive solution
can be seen in Fig.~\ref{fig3final}(b), where the 
acceleration of the maximum maximorum (solid blue line), of the first minima (dashed blue line), and of the zero-levels (magenta dot-dashed line) are shown. For the maximum maximorum, the acceleration is diminished, whereas for the first minima (and other maxima and minima) the accelerating properties remain the same. Furthermore, the modulated density solution \eqref{airy2} diffuses slower than solution \eqref{airy1}, and different to Gaussian solutions. In Fig.~\ref{fig4},  we use
density \eqref{airy2} to calculate $\sqrt{t}\, \phi$ evaluated in its maximum maximorum (in blue solid line). This approaches to $\sim\sqrt{t}$ as time grows. Similarly,  density \eqref{airy2} evaluated in its first minima (blue dashed line) shows a diffusion 
rate of the order $\sim{t^2}$ (in black dotted line) as time grows.

\section{Three--dimensional General accelerating  diffusion}

The above solution can be generalized to a general three-dimensional system, described by Eq.~\eqref{difGenera} for a general constant diffusion tensor, with components $D_{ij}\neq 0$, for $i,j=1,2,3$.
Employing
the same mathematical tools from previous section, it is possible to show that an accelerating solution of Eq.~\eqref{difGenera} is
\begin{eqnarray}
\phi(t,x,y,z)&=&{\mbox{Ai}}\left(k_1 x+k_2 y+k_3 z+ \gamma^2 t^2 \right)\nonumber\\
&&\times\exp\left(\gamma t\left(k_1 x+k_2 y+k_3 z\right)+ \frac{2}{3}\gamma^3 t^3 \right)\, ,\nonumber\\
&&
\label{airy3d}
\end{eqnarray}
with $\gamma=D_{ij}k_ik_j$. We emphasize that this solution is for a general constant diffusion with diffusive density
in every possible direction.
Notice that  general solution \eqref{airy3d} is equivalent to one--dimensional solution \eqref{airy1} by performing the change of variables $x'=(k_1 x+k_2 y+k_3 z)/k$,
with $k=(k_1^2+k_2^2+k_3^2)^{1/2}$.

 On the other hand, in the particular case of a constant diffusion tensor with the form $D_{ij}=D_i\delta_{ij}$ (with a delta function $\delta_{ij}$) the following accelerating density is a different  solution to Eq.~\eqref{difGenera}
\begin{eqnarray}
\phi(t,x,y,z)&=&{\mbox{Ai}}\left(\kappa_1 x+ \varrho_1^2 t^2 \right){\mbox{Ai}}\left(\kappa_2 y+ \varrho_2^2 t^2 \right)\nonumber\\
&&\times{\mbox{Ai}}\left(\kappa_3 z+ \varrho_3^2 t^2 \right)\nonumber\\
&&\times\exp\left(t(\kappa_1\varrho_1 x+\kappa_2\varrho_2 y+\kappa_3\varrho_3 z)+\varsigma t^3\right)\, ,\nonumber\\
&&
\end{eqnarray}
with $\varrho_i=\kappa_i^2 D_i$,
and $\varsigma=2(\varrho_1^3+\varrho_2^3+\varrho_3^3)/3$.

Lastly, to construct a modulated accelerating version of the solutions above, the same scheme 
followed in previous sections may be used.

\section{Discussion}

The accelerating diffusion processes presented above have
 interesting  features that depend on the initial conditions of the system. They are the Wick rotated analogues of the accelerating solutions that appear  in quantum mechanics in theory \cite{hojmanasenjo} and experiments \cite{Broky}.  In this sense, this accelerating diffusion phenomena  can be harnessed  in new processes where now the way it diffuses can be directed in space by proper choices of the free parameters in the solution.

Although other accelerating diffusion processes which propagate depending on initial conditions do exist, such as what happens in reaction-diffusion equations with 
faster diffusion and long distance dispersal \cite{garnier,hamel,dstan,rking}, the locally accelerating system presented here satisfy an equation which does not present non-linearities. It belongs to the same linear diffusive phenomena as Gaussian densities.

The above accelerating behavior corresponds to local properties of the solutions \eqref{airy1}  and \eqref{airy2}. They display different kind of accelerations that involve different dynamics than the ones of Gaussian solutions. However, their global features are analogous to their Gaussian counterparts. For instance, the average position of densities \eqref{airy1} and \eqref{airy2} is $\langle x\rangle=\int_{-\infty}^\infty x\,  \phi(x,t) dx=\mbox{constant}$, that depends only on $\eta$ (for $\eta=0$, then 
$\langle x\rangle=0$), and thus the global solution behaves in the same way as Gaussian ones do, with $d\langle x\rangle/dt=0$. 
On the other hand, both mean square distance of densities 
\eqref{airy1} and \eqref{airy2} result to be
\begin{equation}
\langle x^2\rangle\sim t\, ,
\end{equation} 
as Gaussian solutions. Therefore, the accelerating densities have an analogue global behavior to standard solutions of diffusion.

Finally, we expect to have other possible accelerating solutions of this kind for the nonlinear version of the diffusion equation. These endeavors are left for future work.



\begin{thebibliography}{17}

\bibitem{crank} J. Crank, {\it The Mathematics of Diffusion} (Oxford University Press, 1975).
\bibitem{blackscholes} F. Black and M. Scholes,  Jour. Political Economy {\bf 81}, 637 (1973). 
\bibitem{bioph} D. B. Chang {\it et al.}, Jour. Theo. Bio. {\bf 50},  285 (1975).
\bibitem{burgers} J. M. Burgers, {\it The Nonlinear Diffusion Equation:
Asymptotic Solutions and Statistical Problems} (Springer Netherlands, 1974).
\bibitem{king1} J. R. King, J. Phys. A: Math. Gen. {\bf 23}, 3681 (1990).
\bibitem{king2} J. R. King, J. Phys. A: Math. Gen. {\bf 24}, 3213 (1991).
\bibitem{chole} F. M. Cholewinski and J. A. Reneke, Elec. Jour. Diff. Eqs., {\bf 2003}, 1 (2003).
\bibitem{metzler} R. Metzler and J. Klafter, Europhys. Lett. {\bf 51}, 492 (2000).
\bibitem{chai} Chai Hok Eab and S. C. Lim, J. Phys. A: Math. Theor. {\bf 45}, 145001 (2012).
\bibitem{said} E. A. Saied and M. M. Hussein, J. Phys. A: Math. Gen. {\bf 27} 4867 (1994).
\bibitem{ander} D. Anderson and M. Lisak, Phys. Rev. A {\bf 22}, 2761 (1980).
\bibitem{james} J. M. Hill, A. J. Avagliano, M. P. Edwards,  IMA Journal of Applied Mathematics {\bf 48}, 283 (1992).





\bibitem{tsekov} R. Tsekov, Phys. Scr. {\bf 83}, 035004 (2011). 

\bibitem{garnier} J. Garnier, SIAM J. Math. Anal. {\bf 43}, 1955 (2011).
\bibitem{dstan} D. Stan and J. L. V\'azquez, SIAM  J. Math. Anal. {\bf 46}, 3241 (2014).
\bibitem{hamel} F. Hamel and L. Roques, J. Diff. Equations {\bf249}, 1726 (2010).
\bibitem{rking} R. King and P. M. McCabe, Proc. R. Soc. A  {\bf 459}, 2529 (2003).

\bibitem{berry} M. V. Berry and N. L. Balazs, Am. J. Phys. {\bf 47}, 264 (1979).
\bibitem{sivi} G. A. Siviloglou and D. N. Christodoulides, Opt. Lett. {\bf 32}, 979 (2007).
\bibitem{lekn} J. Lekner, Eur. J. Phys. {\bf 30} L43 (2009).
\bibitem{hojmanasenjo} S. A. Hojman and F. A. Asenjo, Phys. Lett. A {\bf 384}, 126913 (2020).
\bibitem{Broky} G. A. Siviloglou, J. Broky, A. Dogariu, and D. N. Christodoulides,
Phys. Rev. Lett. {\bf 99}, 213901  (2007).

\end{thebibliography}
\end{document}